\documentclass[11pt, a4paper, notitlepage]{article}

\usepackage{amsmath, latexsym, amsthm, amsfonts, amssymb} 
\usepackage{graphicx} 
\usepackage{epsfig}
\usepackage{varioref}

\theoremstyle{plain}
\newtheorem{thm}{Theorem}[section]

\newtheorem{lem}{Lemma}[section]

\theoremstyle{definition} 

\makeindex

\newcommand{\beao}{\begin{eqnarray*}}
\newcommand{\eeao}{\end{eqnarray*}\noindent}

\newcommand{\beam}{\begin{eqnarray}}
\newcommand{\eeam}{\end{eqnarray}\noindent}

\newcommand{\beqq}{\begin{equation}}
\newcommand{\eeqq}{\end{equation}\noindent}

\newcommand{\bce}{\begin{center}}
\newcommand{\ece}{\end{center}}

\newcommand{\barr}{\begin{array}}
\newcommand{\earr}{\end{array}}

\begin{document}

\title {{\em Preliminary Version} \\ Evidential Value in ANOVA-Regression Results in Scientific Integrity Studies}
\author{
Chris A.J. Klaassen,\\
Korteweg-de Vries Institute for Mathematics \\
University of Amsterdam\\
P.O. Box 94248, 1090 GE Amsterdam, The Netherlands\\
 email: c.a.j.klaassen@uva.nl}

\maketitle

\begin{abstract}
Some scientific publications are under suspicion of fabrication of
data. Since humans are bad random number generators, there might
be some evidential value in favor of fabrication in the
statistical results as presented in such papers. In case of
ANOVA-Regression studies we present the evidential value of the
results of such a study in favor of the hypothesis of a dependence
structure in the underlying data, which indicates fabrication,
versus the hypothesis of independence, which is the ANOVA model
assumption. Applications of this approach are also presented.
\end{abstract}

\section{Introduction}\label{I}

Consider a publication based on an empirical study in which the
data are analyzed by an ANOVA-Regression model, as presented in
Section \ref{MFD}. Assume that the study is under suspicion of
data fabrication and that the underlying data are not available.
Based on just the publication itself it has to be decided if the
suspicion is justified. To this end we implement an idea of
Simonsohn (2012), which states that typically when data are
fabricated, the model fit is {\em too good to be true}, and we
apply the by now standard approach in Forensic Statistics, which
is a Bayesian one, as described in Section \ref{EVSIS}. This
approach yields the so-called evidential value of the publication
in favor of the hypothesis of a dependence structure in the
underlying data, which indicates fabrication, versus the
hypothesis of independence, which is the ANOVA model assumption.
In Section \ref{EVFD}, in particular in Theorem \ref{V} we
describe how this evidential value may be computed, or at least be
bounded from below and from above. This Theorem is proved in the
Appendix. We apply our approach to F\"orster and Denzler
(2012) in Section \ref{A}. Some notes on the interpretation of
evidential value are presented in Section \ref{IEV}. Our approach is similar to the one in Klaassen (2013).

\section{Evidential Value in Scientific Integrity Studies}\label{EVSIS}

The hypothesis $H_F$ of fabrication of data has been put forward
about a scientific publication. The author claims the hypothesis
$H_I$ of integrity of the data holds. A Committee on Scientific
Integrity has to decide in favor of $H_F$ or $H_I.$ In line with
the so-called {\em Bayesian Paradigm of Forensic Statistics} the
Committee has to construct a prior opinion about $H_F$ and $H_I,$
i.e. before studying the evidence $E,$ namely the paper itself and
other evidence. This prior opinion is formulated in terms of the
prior odds in favor of the hypothesis of fabrication, namely
$$ P(H_F)\,/\,P(H_I).$$
Subsequently experts have to determine the probability that $E$
occurs under both the hypothesis $H_I$ that the data have been
collected in the scientifically right way and the hypothesis $H_F$
that the data have been manipulated or fabricated. The ratio of
these probabilities
$$P(E\,|\,H_F)\,/\,P(E\,|\,H_I)$$ is called the likelihood ratio.
Multiplying the prior odds and the likelihood ratio the Committee
obtains the so-called posterior odds in favor of the hypothesis of
fabrication $$P(H_F\,|\,E)\,/\,P(H_I\,|\,E),$$ i.e., the odds in
favor of $H_F$ after having seen the evidence. The Committee has
to base its decision on these posterior odds. In summary, the {\em
Bayesian Paradigm of Scientific Integrity Studies} reads as
follows
\begin{equation}\label{BayesParadigm}
 \underbrace{\frac{P(H_F)}{P(H_I)}}_{prior\, odds}\ \
\overbrace{\frac{P(E\,|\,H_F)}{P(E\,|\,H_I)}}^{likelihood\,
ratio}\, =\
\underbrace{\frac{P(H_F\,|\,E)}{P(H_I\,|\,E)}}_{posterior\,
odds}\,.
\end{equation}\\
Since the likelihood ratio in (\ref{BayesParadigm}) may be
interpreted as the weight that the evidence should have in the
decision of the Committee, it is called the evidential value in
favor of the hypothesis of fabrication (versus the hypothesis of
integrity).

The evidence $E$ is viewed here as a realization of a random
mechanism, both under $H_F$ and $H_I.$ In case this random
mechanism produces outcomes via probability density functions
$f(E\,|\,H_F)$ and $f(E\,|\,H_I),$ the probabilities in the
likelihood ratio or evidential value are replaced by the
corresponding probability density functions, resulting in
\begin{equation}\label{BayesParadigmdensity}
 \underbrace{\frac{P(H_F)}{P(H_I)}}_{prior\, odds}\ \
\overbrace{\frac{f(E\,|\,H_F)}{f(E\,|\,H_I)}}^{evidential\,
value}\, =\
\underbrace{\frac{P(H_F\,|\,E)}{P(H_I\,|\,E)}}_{posterior\,
odds}\,.
\end{equation}\\

\section{Modelling Fabrication of Data Underlying a Specific Type
of ANOVA-Regression Studies}\label{MFD}

In one-way Analysis of Variance the basic assumption is that all
observations may be viewed as realizations of independent normally
distributed random variables with means that depend on the values
of some categorical covariate. Let this categorical covariate take
three values only, and let the number of observations for each of
the three cells be the same, namely $n.$ The random variables
denoting the observations are then
\begin{equation}\label{model}
X_{ij}= \mu_i +\varepsilon_{ij}\,,\quad i=1, 2, 3,\, j=1,\dots, n.
\end{equation}
The cell means $\mu_i$ are unknown real numbers, and the
measurement errors $\varepsilon_{ij}$ are independent, normally
distributed random variables with mean 0 and variance
$\sigma_i^2,\ i=1, 2, 3.$ Contrary to the standard assumption in
ANOVA we assume here that the variances may depend on the
covariate. In our ANOVA-regression model there exist constants
$\alpha$ and $\beta,$ such that
\begin{equation}
\mu_i =\alpha + \beta i,\quad i=1,2,3.
\end{equation}
Actually, this is equivalent to the following restriction on the
means $\mu_i$
\begin{equation}\label{regression}
\mu_1-2\mu_2+\mu_3=0.
\end{equation}
If authors are fiddling around with data and are fabricating and
falsifying data, they tend to underestimate the variation that the
data should show due to the randomness within the model. Within
the framework of the above ANOVA-regression case, we model this by
introducing dependence between the normal random variables
$\varepsilon_{ij},$ which represent the measurement errors.
Actually, we assume that the measurement errors in any cell may have nonzero
correlation to the corresponding measurement errors in
the other cells. More precisely formulated, we assume that the
correlations between the random variables $\varepsilon_{ij}$ no
longer all vanish, but satisfy
\begin{equation}\label{dependence}
\rho(\varepsilon_{1j}, \varepsilon_{2j})= \rho_3,\
\rho(\varepsilon_{1j}, \varepsilon_{3j})=\rho_2,\
\rho(\varepsilon_{2j}, \varepsilon_{3j})= \rho_1,\quad
j=1,\dots,n,
\end{equation}
with all other correlations still being equal to 0. Because we restrict attention to multivariate normal densities in the sequel,
we exclude $|\rho_i|=1,$ so we assume $-1 < \rho_i <
1,\, i=1, 2, 3.$ We note that under the standard assumptions of
ANOVA $\rho_i=0$ holds. Furthermore, we note that within cells
observations may be renumbered in order to get the structure
(\ref{dependence}). Nevertheless, we still assume (\ref{model}) to
hold and the measurement errors to be normally distributed with
mean 0 and variance $\sigma_i^2,\ i=1, 2, 3.$
Since the covariance matrix of the $\varepsilon_{i1}$s has to be positive
semidefinite, the determinant
\begin{eqnarray}\label{determinant}
\left|
\begin{array}{ccc}
\sigma_1^2 & \sigma_1 \sigma_2 \rho_3 & \sigma_1 \sigma_3 \rho_2
\\
\sigma_1 \sigma_2 \rho_3 & \sigma_2^2 & \sigma_2 \sigma_3 \rho_1
\\
\sigma_1 \sigma_3 \rho_2 & \sigma_2 \sigma_3 \rho_1 & \sigma_3^2
\end{array} \right|
= \sigma_1^2 \sigma_2^2 \sigma_3^2 \left(1-\rho_1^2 -\rho_2^2
-\rho_3^2 +2\rho_1 \rho_2 \rho_3 \right)
\end{eqnarray}
has to be nonnegative, and hence we have the side condition
$1-\rho_1^2 -\rho_2^2 -\rho_3^2 +2\rho_1 \rho_2 \rho_3 \geq 0$ on the
$\rho$'s. Again, for technical reasons we prefer to work with multivariate normal densities and hence we shall assume $1-\rho_1^2 -\rho_2^2 -\rho_3^2 +2\rho_1 \rho_2 \rho_3 > 0.$

A way in which fabrication of measurement errors may take place is
by copying some of them with an additional multiplication and
addition or subtraction.
 This might be modelled as follows. Let
$U_j\,,\, j=1, \dots, n,$ and $V_{ij}\,,\, i=1, 2, 3,\ j=1, \dots,
n,$ be independent and identically distributed standard normal
random variables. Independent of these, let the random indicators
$\Delta_{ij}\,,\, i=1, 2, 3,\ j=1, \dots, n,$ be independent
Bernoulli random variables with $P(\Delta_{1j}=1)=\sqrt{\rho_2
\rho_3/\rho_1}, P(\Delta_{2j}=1)=\sqrt{\rho_1 \rho_3/\rho_2},
P(\Delta_{3j}=1)=\sqrt{\rho_1 \rho_2/\rho_3},$ and
$P(\Delta_{1j}=0)=1-P(\Delta_{1j}=1),$ etc. Then
\begin{equation}\label{structure}
\epsilon_{ij}=\sigma_i \left(\Delta_{ij} U_j + (1-\Delta_{ij})
V_{ij}\right)
,\quad i=1, \dots, I,\ j=1, \dots, n,
\end{equation}
satisfy (\ref{model}). Note that we have
$\varepsilon_{1j}/\sigma_1= \varepsilon_{2j}/\sigma_2= U_j$ with
probability $\rho_3$ then, and since analogous relations hold for
the two other combinations, the measurement errors satisfy
(\ref{dependence}).

Finally, we note that (\ref{dependence}) is just one possible way
to model dependence, and that the actual way in which fabrication
has been implemented, might lead to quite different dependence
structures. However, this model will come close to some types of
fabrication and falsification.

\section{Evidential Value for Fabrication of Data Underlying an ANOVA-Regression Study}\label{EVFD}

Consider a study in a scientific research paper. The data
underlying this study are analyzed by the one-way layout ANOVA
model (\ref{model}) of the preceding section and as results the
sample means and sample standard deviations of the three cells are presented. The
underlying data themselves are not published and are not
available. According to the theory as developed in the research paper the
linear regression condition (\ref{regression}) holds.

There are two hypotheses to be formulated about the data
underlying this ANOVA-regression study. The hypothesis $H_F$ of
fabrication of the data underlying the results presented in the
paper, is that $\rho_i \neq 0$ holds for at least one $i,\, i=1, 2,
3.$ The other hypothesis $H_I$ represents the situation that data
have been collected according to (\ref{model}) with independent
$X_{ij},$ i.e., $\rho_1=\rho_2=\rho_3=0.$ We want to determine the
evidential value of the published results of the ANOVA-regression
study, i.e., of the sample means and sample variances for the three cells,
in favor of the hypothesis $H_F$ versus $H_I.$

To this end we first note that the sample means in the cells,
\begin{equation}\label{sampleaverage}
X_i=\frac 1n \sum_{j=1}^n X_{ij},\quad i=1, 2, 3,
\end{equation}
have a joint trivariate normal distribution. Actually, the
dependence structure (\ref{dependence}) implies
\begin{equation}\label{multinormal}
\begin{pmatrix}
X_1 \cr X_2 \cr X_3 \cr
\end{pmatrix}
\sim {\cal N} \left(
\begin{pmatrix}
\mu_1 \cr \mu_2 \cr \mu_3 \cr
\end{pmatrix}
,\, \mbox{$n^{-1}$}
\begin{pmatrix}
\sigma_1^2 & \sigma_1 \sigma_2 \rho_3 & \sigma_1 \sigma_3 \rho_2 \cr
\sigma_1 \sigma_2 \rho_3 & \sigma_2^2 & \sigma_2 \sigma_3 \rho_1
\cr \sigma_1 \sigma_3 \rho_2 & \sigma_2 \sigma_3 \rho_1 &
\sigma_3^2 \cr
\end{pmatrix}
\right).
\end{equation}
In stead of assuming normally distributed errors satisfying
(\ref{dependence}), we could have started right away from
(\ref{multinormal}). This is a much weaker condition that in
practice is more likely to be satisfied approximately in view of
the central limit theorem.

By $X$ we denote the column 3-vector with components $X_1, X_2,$
and $X_3.$ Let these components be uncorrelated and let $A$ be a nonsingular $3\times 3$-matrix such that
the components $Y_1, Y_2,$ and $Y_3$ of $Y=AX$ are uncorrelated as well,
and hence by the normality assumption independent. The first row
of $A$ is chosen to be $(1,-2,1),$ which entails $Y_1=X_1
-2X_2+X_3.$ The two other rows depend on the values of the
parameters in the covariance matrix, but not on the value of the
3-vector $\mu=(\mu_1,\mu_2,\mu_3)^T.$ So, for inference about
$\mu$ the vector $Y$ is equivalent to $X.$ Let
$\nu=(\nu_1,\nu_2,\nu_3)^T=EY$ be the expectation of $Y.$ Because
of the nonsingularity of $A$ there does not exist a linear
combination of $\nu_2$ and $\nu_3$ that equals $\nu_1.$ By the
independence of the components of $Y$ this implies that the first
component $Y_1$ is a sufficient statistic for its expectation
$\nu_1=\mu_1-2\mu_2+\mu_3,$ which according to the theory as
claimed by the paper under study vanishes, as in
(\ref{regression}). This means that all information about $\nu_1$
contained in the independent sample means $X_1, X_2,$ and $X_3,$ is contained in $Y_1=X_1
-2X_2+X_3.$ Therefore we will base our evidential value on this
statistic, which we will rename as $Z=X_1 -2X_2+X_3.$

First we note that under the linear regression assumption (\ref{regression})
we have
\begin{eqnarray}\label{generalmodel}
\lefteqn{{\sqrt n}\left(X_1 -2X_2+X_3 \right)= {\sqrt n}\,Z \sim {\cal N} \left( 0, \sigma_Z^2 \right),}\\
&& \sigma_Z^2 = \sigma_1^2 +
4 \sigma_2^2 + \sigma_3^2 -4 \sigma_1 \sigma_2 \rho_3 + 2 \sigma_1
\sigma_3 \rho_2 - 4 \sigma_2 \sigma_3 \rho_1 . \nonumber
\end{eqnarray}
This normal distribution depends on the parameters
$\rho_1,\rho_2,\rho_3, \sigma_1^2, \sigma_2^2, \sigma_3^2,$ with
$-1 < \rho_i <1,\, 1-\rho_1^2 -\rho_2^2 -\rho_3^2 +2\rho_1 \rho_2 \rho_3 > 0,\, 0 < \sigma_i,\, i=1,2,3.$
In the studies we consider, only realizations $x_i$ of
the cell means $X_i$ and estimates $s_i^2$ of the cell variances
$\sigma_i^2$ are given, $i=1, 2, 3.$

Let us denote the density of
$Z=X_1-2X_2+X_3$ at $z=x_1-2x_2+x_3$ with $\sigma_i$ replaced by
$s_i>0$ by
\begin{eqnarray}\label{density}
f_n(z; \rho_1, \rho_2, \rho_3)= \sqrt{\frac n{2\pi}}\ \frac
1{s(\rho_1,\rho_2,\rho_3)}\ \exp\left(-\frac {nz^2}{2s^2(\rho_1,\rho_2,\rho_3)}\right),\\
s(\rho_1,\rho_2,\rho_3)=\left( s_1^2 + 4 s_2^2 + s_3^2 -4 s_1 s_2
\rho_3 + 2 s_1 s_3 \rho_2 - 4 s_2 s_3 \rho_1 \right)^{1/2}. \nonumber
\end{eqnarray}
We will base our evidential value on this density, viewing $s_1, s_2,$ and $s_3$ as given.

The hypothesis $H_I$
of proper data corresponds to $\rho_1=\rho_2=\rho_3 = 0.$ We shall let the hypothesis $H_F$ of fabrication of the data correspond
to nonzero correlation between at least two sample means, such that
\begin{equation}\label{boundedvaraine}
s(\rho_1,\rho_2,\rho_3) \leq s(0,0,0)
\end{equation}
holds. This means that we restrict $H_F$ by the condition that ${\sqrt n}Z = {\sqrt n}(X_1 -2 X_2 + X_3)$ has a(n estimated) variance that equals at most the (estimated) variance under independence, $H_I.$ This restriction is in line with our presumption that people when fabricating data tend to underestimate variation.
The evidential value
$$\frac{f(E\,|\,H_F)}{f(E\,|\,H_I)}$$ from
(\ref{BayesParadigmdensity}) in favor of $H_F$ versus $H_I$
becomes in this case (cf. Zhang (2009), Bickel (2012))
\begin{equation}\label{evidentialvalue1}
\mathbb V = \frac{\sup_{-1< \rho_i<1,\, 1-\rho_1^2 -\rho_2^2
-\rho_3^2 +2\rho_1 \rho_2 \rho_3 > 0,\,s(\rho_1,\rho_2,\rho_3) \leq s(0,0,0)} f_n(z; \rho_1, \rho_2,
\rho_3)} {f_n(z;0,0,0)}.
\end{equation}
This evidential value may be computed with the help of the
following Theorem.
\begin{thm}\label{V}
Within the general model (\ref{generalmodel}) and with the notation (\ref{density}), define
$$s_L^2=\inf_{-1< \rho_i<1,\, 1-\rho_1^2 -\rho_2^2
-\rho_3^2 +2\rho_1 \rho_2 \rho_3 > 0,\, s(\rho_1,\rho_2,\rho_3) \leq s(0,0,0)}s^2(\rho_1,\rho_2,\rho_3)$$
 and $${\tilde s}_L^2= \min\left\{\left(2s_2 -(s_1 + s_3)\right)^2, \left(2s_2 - \sqrt{s_1^2+s_3^2}\right)^2 \right\},$$
and write
$s_0^2=s^2 (0,0,0)=s_1^2+4s_2^2+s_3^2.$
Then
\begin{equation}\label{inequalities}
s_L^2 \leq {\tilde s}_L^2 \leq s_0^2
\end{equation}
holds. Furthermore, we have:
\begin{itemize}
\item
If $${\tilde s}_L^2 \leq nz^2 \leq s_0^2$$ holds, then the
evidential value from (\ref{evidentialvalue1}) becomes
\begin{equation}\label{value1}
\mathbb V = \frac{s_0}{\sqrt{nz^2}} \exp\left\{-\frac 12 nz^2
\left[\frac 1{nz^2}-\frac 1{s_0^2} \right] \right\}\geq 1.
\end{equation}
\item
If $$nz^2 \leq {\tilde s}_L^2$$ holds, then the evidential value from
(\ref{evidentialvalue1}) satisfies
\begin{equation}\label{value3}
\mathbb V \geq \frac{s_0}{{\tilde s}_L} \exp\left\{-\frac 12 nz^2
\left[\frac 1{{\tilde s}_L^2} -\frac 1{s_0^2}\right]\right\}\geq 1
\end{equation}
and equals at most the left hand side of inequality (\ref{value1}).
\item
If $$s_0^2 \leq nz^2$$ holds, then the evidential value from
(\ref{evidentialvalue1}) becomes $\mathbb V =  1.$
\end{itemize}
\end{thm}
The proof is given in the Appendix.

\section{Application}\label{A}

A complaint has been filed about the scientific integrity of F\"orster and Denzler
(2012). This paper contains 12 studies modelled as in Section
\ref{MFD}. The three cell means, $x_1, x_2, x_3,$ for each study
have been given in the paper. The sample standard deviations for these cells,
$s_1, s_2, s_3,$ have been provided by the authors to the
investigator who filed the complaint. These data and the
corresponding evidential values are given in Table 1.

\begin{table}
\caption{The evidential values of 12 studies from F\"orster and Denzler (2012).}
\medskip
\centering
\begin{tabular}{|l|l|l|l|l|}\hline
Study      & $n$     &  $x_1, x_2, x_3$  & $s_1, s_2, s_3$ &
$\mathbb V$ \\\hline
1 & 20 & 2.47, 3.04, 3.68 & 1.21, 0.72, 0.68 & 3.92 \\
2 & 20 & 2.51, 2.95, 3.35 & 0.71, 0.49, 0.64 & 4.68 \\
3 & 20 & 2.40, 2.90, 3.45 & 0.86, 0.51, 0.80 & 4.26 \\
4 & 20 & 2.41, 2.98, 3.64 & 1.07, 0.51, 0.95 & 2.72 \\
5 & 20 & 2.14, 2.82, 3.41 & 1.20, 0.78, 0.71 & 3.21 \\
6 & 20 & 3.19, 4.01, 4.79 & 1.07, 1.21, 0.82 & 4.95--9.41 \\
\hline
7 & 20 & 2.63, 3.73, 4.73 & 1.49, 1.21, 1.55 & 4.43 \\
8 & 20 & 2.87, 3.83, 4.79 & 1.24, 1.09, 1.53 & 13.95--$\infty$ \\
9a & 20 & 2.35, 3.66, 4.76 & 1.01, 1.19, 1.71 & 2.10 \\
9b & 15 & 2.55, 3.72, 4.78 & 1.16, 1.00, 1.47 & 3.95 \\
10a & 20 & 2.66, 3.69, 4.81 & 1.21, 1.30, 1.54 & 4.94 \\
10b & 15 & 2.42, 3.73, 5.02 & 0.82, 1.28, 1.45 & 10.17--23.92 \\
\hline
\end{tabular}\\
\end{table}

To interpret these evidential values it is useful to consider also
the evidential values that are obtained for similar publications
in the same field as collected in the complaint; see Table 2.

\begin{table}
\caption{Evidential values of 21 studies from the social psychology literature.}
\medskip
\centering
\begin{tabular}{|l|l|l|l|l|}
\hline
Study      & $n$     &  $x_1, x_2, x_3$  & $s_1, s_2, s_3$ &
$\mathbb V$ \\\hline
Hagtvedt-l & 141/6 & 4.39, 3.97, 3.84 & 0.76, 1.26, 1.14 & 1.40 \\
Hagtvedt-2 & 141/6 & 3.22, 3.84, 4.11 & 0.98, 1.02, 1.46 & 1.17 \\
Hunt & 75/3 & 1.48, 1.04, 1.04 & 0.82, 0.68, 0.68 & 1 \\
Jia & 132/3 & 1.09, 0.70, 0.59 & 0.89, 0.69, 0.62 & 1 \\
Kanten-l & 269/6 & 3.29, 3.14, 2.66 & 1.11, 0.94, 0.71 & 1.001 \\
\hline
Kanten-2 & 269/6 & 3.02, 2.99, 2.85 & 0.80, 0.84, 0.70 & 1.75 \\
Lerouge-l & 63/3 & 4.24, 2.48, 2.14 & 1.51, 2.16, 2.13 & 1 \\
Lerouge-2 & 63/3 & 2.95, 2.81, 2.62 & 2.44, 1.81, 2.25 &
12.23--13.01 \\
Lerouge-3 & 54/3 & 4.90, 3.31, 2.79 & 2.22, 2.09, 1.66 & 1.01 \\
Lerouge-4 & 54/3 & 3.69, 2.67, 2.50 & 2.78, 2.51, 1.66 & 1.21 \\
\hline
Malkoc & 521/3 & 4.72, 5.36, 6.19 & 4.96, 9.08, 10.58 &
5.26--5.27 \\
Polman & 65/3 & 4.69, 3.50, 2.91 & 2.37, 2.09, 2.42 & 1.34 \\
Rook-l & 168/6 & 6.22, 6.13, 4.73 & 3.05, 2.19, 1.95 & 1 \\
Rook-2 & 168/6 & 5.39, 5.22, 4.61 & 2.14, 2.58, 2.28 & 1.69 \\
Smith-l & 73/3 & 4.38, 4.26, 3.55 & 1.53, 1.36, 1.07 & 1.01 \\
\hline
Smith-2 & 76/3 & 14.83, 12.69, 11.88 & 4.62, 4.95, 4.75 & 1.26 \\
Smith-3 & 113/3 & 0.42, 0.53, 0.56 & 0.20, 0.19, 0.19 & 1\\
Smith-4 & 140/3 & 4.70, 7.90, 11.80 & 7.40, 11.40, 20.40 & 4.04 \\
Smith-5 & 125/3 & 14.52, 13.43, 12.85 & 2.81, 3.27, 3.94 & 1.63 \\
Smith-6 & 97/3 & 10.85, 8.64, 8.32 & 5.07, 3.61, 4.17 & 1 \\
Smith-7 & 144/3 & 4.64, 4.84, 5.49 & 1.30, 1.56, 1.28 & 1.02 \\
\hline
\end{tabular}
\end{table}


We notice that these evidential values from literature are all
below 2, say, except for Lerouge-2, Malkoc, and Smith-4. In
contrast all evidential values from F\"orster and Denzler (2012) are above 2. From Table 2 one might estimate the probability $P_{H_I}({\mathbb V} \geq 2)$ under the hypothesis $H_I$ that the evidential value will equal at least 2, as 3/21 = 1/7. This would imply that the probability that 12 studies will have an evidential value of at least 2, as occurs in Table 1, equals approximately $(1/7)^{12} \approx 7.2 \times 10^{-11}.$

On the other hand, Theorem \ref{V} shows that ${\mathbb V} \geq v >1$ implies
\begin{equation}\label{application}
\frac{s_0}{\sqrt{nz^2}} \exp\left\{\frac 12
\left[\frac {nz^2}{s_0^2}- 1 \right] \right\}\geq v,\quad \frac{nz^2}{s_0^2} \leq 1.
\end{equation}
For $v=2$ some computation shows that this means approximately
\begin{equation}\label{ratio}
- 0.3191 \leq \frac {{\sqrt n}(x_1 - 2x_2 + x_3)}{\sqrt{s_1^2 + 4s_2^2 + s_3^2}} \leq 0.3191.
\end{equation}
Viewing the sample means $x_i$ and sample variances $s_i^2$ as random, we see that the ratio in (\ref{ratio}) has approximately a standard normal distribution under the hypothesis $H_I,$ which implies $P_{H_I}({\mathbb V} \geq 2) \approx 0.2504.$ This is much more than the 1/7 estimated from Table 2, but it still shows that the probability that 12 studies will have an evidential value of at least 2, as in Table 1, equals approximately $(0.2504)^{12} \approx 6.1 \times 10^{-8}.$

\section{Interpreting Evidential Value}\label{IEV}

With the evidential value $\mathbb V$ defined as in
(\ref{evidentialvalue1}) through (\ref{value3}) the Bayesian
paradigm for criminal court cases (\ref{BayesParadigmdensity})
becomes
\begin{equation}\label{BayesParadigmV}
 \underbrace{\frac{P(H_F)}{P(H_I)}}_{prior\, odds}\ \
\overbrace{\mathbb V}^{evidential\, value}\, =\
\underbrace{\frac{P(H_F\,|\,E)}{P(H_I\,|\,E)}}_{posterior\,
odds}\,.
\end{equation}
An important principle in criminal court cases is `in dubio pro
reo', which means that in case of doubt the accused is favored. In
science one might argue that the leading principle should be `in
dubio pro scientia', which should mean that in case of doubt a
publication should be withdrawn. Within the framework of this
paper this would imply that if the posterior odds in favor of
hypothesis $H_F$ of fabrication equal at least 1, then the
conclusion should be that $H_F$ is true. So an ANOVA-regression
study for which
\begin{equation}\label{BayesParadigmV2}
 \underbrace{\frac{P(H_F)}{P(H_I)}}_{prior\, odds}\ \
\overbrace{\mathbb V}^{evidential\, value}\, =\
\underbrace{\frac{P(H_F\,|\,E)}{P(H_I\,|\,E)}}_{posterior\, odds}
> 1
\end{equation}
holds, should be rejected and disqualified scientifically. Keeping
this in mind one wonders what a reasonable choice of the prior
odds would be.

In criminal court cases the choice of prior odds is left to the
judge, and the evidential value has to be determined by the
forensic expert.

We conclude with some notes.
\begin{itemize}
\item
ANOVA studies are based on the assumption of normality. Often this
assumption is not satisfied, but the technique is still applied.
In view of the central limit theorem cell means like in our basic
model (\ref{multinormal}) behave approximately like (jointly
multivariate) normal random variables.
\item
Note that Theorem \ref{V} implies
$$\mathbb V \geq 1.$$
Consequently, within this framework there does not exist
exculpatory evidence. This is reasonable since bad science cannot
be compensated by very good science. It should be very good
anyway.
\item
When a paper contains more than one study based on independent
data, then the evidential values of these studies can and may be
combined into an overall evidential value by multiplication in
order to determine the validity of the whole paper.
\item
We have modelled the hypothesis of data fabrication via
(\ref{multinormal}). However, other, nonnormal multivariate
distributions might model fabrication better. Consequently, higher
evidential values might be possible.
\item
The discussion at the end of Section \ref{A}, in particular the argument involving (\ref{application}) and (\ref{ratio}), shows that the approach of evidential value $\mathbb V$ is just a way to interpret the value of the statistic
\begin{equation}\label{statistic}
Z_{\mathbb V} =\frac{{\sqrt n}(X_1 - 2X_2 + X_3)}{\sqrt{S_1^2 + 4 S_2^2 + S_3^2 }},
\end{equation}
where $S_i^2$ is the sample variance viewed as a random variable. Note that $Z_{\mathbb V}$ has a standard normal distribution approximately. When it takes on a very small (absolute) value or small (absolute) values repeatedly, the suggestion is raised that data have been manipulated; note that if (\ref{regression}) does not hold, $\mid Z_{\mathbb V} \mid$ will take on large values for $n$ large; see the note below.
The investigator who filed the complaint, based his argument on the ANOVA framework, tested the hypothesis (\ref{regression}) by rejecting it for large (absolute) values of the statistic
\begin{equation}\label{statistic2}
Z_C =\frac{{\sqrt n}(X_1 - 2X_2 + X_3)}{\sqrt{2\left(S_1^2 + S_2^2 + S_3^2 \right) }},
\end{equation}
and noted that the $p$-values corresponding to the studies from Table 1 are suspiciously close to 1, in contrast to those from Table 2.

Note that both $Z_{\mathbb V}$ and $Z_C$ have a normal distribution asymptotically as $n \to \infty$ under (\ref{regression}) with mean 0 and variance 1 and $(\sigma_1^2 + 4 \sigma_2^2 + \sigma_3^2)/[2(\sigma_1^2 + \sigma_2^2 + \sigma_3^2)] \in (1/2, 2),$ respectively. Under $\sigma_1 = \sigma_2 = \sigma_3$ they are both standard normal asymptotically.
\item
By asymptotic theory $Z_{\mathbb V}$ from (\ref{statistic}) has a normal distribution approximately, with mean $\sqrt{n}(\mu_1-2\mu_2+\mu_3)/\sqrt{\sigma_1^2+4\sigma_2^2 + \sigma_3^2}$ and variance 1. Since normal densities are unimodal and symmetric around their mean, this implies that $P\left( -v \leq Z_{\mathbb V} \leq v \right),\ v>0,$ attains its maximum value under (\ref{regression}), at least approximately.
This observation supports the heuristic that $Z_{\mathbb V}$ discerns between $\mu_1-2\mu_2+\mu_3=0$ and $\mu_1-2\mu_2+\mu_3 \neq 0.$ A similar observation holds for $Z_C.$
\item
Since $Z_{\mathbb V}$ and $Z_C$ are are quite similar statistics, the difference between the approach in the present paper and the approach of the compliant is basically the difference between a Bayesian and a frequentist approach. These are just two methods to interpret the data and they point in the same direction, typically.
\end{itemize}

\appendix

\section{Appendix: Proof}

Here we present a proof of Theorem \ref{V}.
In view of
\begin{equation}\label{limits}
s^2(1,1,1)= s_0^2 - 4s_1 s_2 +2s_1 s_3 -4s_2 s_3 = \left(2s_2-s_1- s_3\right)^2,
\end{equation}
\begin{equation*}
s^2\left( \frac{s_3}{\sqrt{s_1^2+s_3^2}}, 0,
\frac{s_1}{\sqrt{s_1^2+s_3^2}}\right)=s_0^2 - 4s_2
\sqrt{s_1^2+s_3^2} = \left(2s_2 -
\sqrt{s_1^2+s_3^2}\right)^2
\end{equation*}
we arrive at the inequalities of (\ref{inequalities}).
The proof of the theorem is completed by repeated application of the
following lemma.

\begin{lem}\label{maximization}
The function $$x \mapsto \frac 1{\sqrt x}\ e^{- \lambda/x}$$ is
increasing from 0 at 0 to $1/\sqrt{2e\lambda}$ at $x=2\lambda,$
and subsequently decreasing to 0 at $\infty.$ Furthermore, the
function $$ x \mapsto \frac 1{\sqrt x}\ e^{\frac 12 (x-1)}$$
attains its minimum value 1 on $(0,\infty)$ at $x=1.$
\end{lem}
\noindent{\bf Proof}\\
Differentiation yields these results.
\hfill$\Box$\\

\end{document}